\documentclass{article}
\usepackage{spconf,amsmath,graphicx}
\usepackage[ruled,vlined]{algorithm2e}
\usepackage{bm}
\usepackage{subfig}
\usepackage{amsfonts}
\usepackage{cite}
\usepackage{comment}

\DeclareMathOperator{\C}{\mathbb{C}}
\DeclareMathOperator{\E}{\mathbb{E}}

\newcommand{\argmin}{\operatornamewithlimits{argmin}}

\title{Spatial-temporal switching estimators for imaging locally concentrated dynamics}
\name{Parisa Karimi$^{*,1}$\thanks{*parisa2@illinois.edu}, Mark Butala$^{2}$, Zhizhen Zhao$^{1}$, Farzad Kamalabadi$^{1}$}
\address{$^{1}$University of Illinois at Urbana-Champain\\
$^{2}$College of information science and electronics engineering, Zhejiang University}
\usepackage{color}

\begin{document}
%
\maketitle
\begin{abstract}
The evolution of images with physics-based dynamics is often spatially localized and nonlinear. A switching linear dynamic system (SLDS) is a natural model under which to pose such problems when the system's evolution randomly switches over the observation interval. Because of the high parameter space dimensionality, efficient and accurate recovery of the underlying state is challenging. The work presented in this paper focuses on the common cases where the dynamic evolution may be adequately modeled as a collection of decoupled, locally concentrated dynamic operators. Patch-based hybrid estimators are proposed for real-time reconstruction of images from noisy measurements given perfect or partial information about the underlying system dynamics. Numerical results demonstrate the effectiveness of the proposed approach for denoising in a realistic data-driven simulation of remotely sensed cloud dynamics. 
\end{abstract}
\section{introduction}
The image formation of time-varying phenomena is commonplace in astronomy \cite{rem2}, remote sensing \cite{galvan12}, and many other disciplines, e.g., \cite{george08}. The challenge is to recover spatial-temporal parameters of a physical phenomenon given noisy, ex situ measurements. Such time-dependent inverse problems can be posed in the framework of a linear dynamic system (LDS), where the state variables represent the volumetric physical parameters of interest which may be reconstructed through Bayesian inference. Oftentimes, the system dynamics traverse or cycle through a sequence of distinct modes, e.g., a sudden, transient shock and then recovery towards a quiescent background state or some pattern caused by an external driver. In such cases, it is natural to augment the LDS with a hidden random variable to represent the system dynamic mode and formulate the problem under a switching LDS (SLDS). A switching Kalman filter (SKF) \cite{skf} is then the optimal approach to jointly estimate the system state and detect the sequence of dynamic modes, but the enormous dimensionality typically involved precludes such an approach.

Physical phenomena often exhibit strong spatial correlations, providing a priori knowledge which may be utilized to dramatically reduce the SLDS parameter dimensionality. For example, cloud dynamics are (approximately) localized in space where the underlying physics are governed by local variables such as humidity and temperature \cite{cloud}. In this work, we harness this prior information to segment the state space as a collection of disjoint convex regions and assume that time evolution at a given point is correlated only with immediate neighbors. 
Under these conditions, we employ a ``divide and conquer'' strategy and develop a patch-based technique to reduce the switching mode cardinality as well as the parameter space dimensionality.

Patch-based algorithms are often used in image and video processing for such tasks as denoising and inpainting \cite{patchim1,patchim2,inp}. For example in video denoising, \cite{bihen1,3d2,3d3}  spatial-temporal patches are used for real-time denoising. To further reduce computation, recursive patch-based denoising algorithms \cite{patch1,patch2,patch3,icip,cvpr} use online clustering to determine the most similar patches per frame, estimate patch cluster dynamics, and track patches by using optical flow.

Generally speaking, conventional patch-based video processing algorithms assume little to no prior information regarding the measurement or the evolution model. Instead, they make use of the inherent redundancy found in natural video frames when collected at high temporal frequency to determine those (possibly disjoint) regions exhibiting a common dynamic mode and track feature motion. In contrast, the physics-based imaging applications considered here must rely on prior information to compensate for measurement deficiencies, e.g., severe noise and sparsity, and feature tracking is challenging or impossible because the hidden object could be diffuse and state dynamics are complicated, e.g., advective flow as opposed to rigid body motion.

The remainder of the paper is organized as follows. Section~\ref{section:model} introduces the SLDS model which captures locally concentrated dynamics. The patch-based estimator is proposed in section \ref{sec:pSKF} and its computational complexity is studied in Section~\ref{section:comp}. Section \ref{section:sim} compares the performance of the proposed filter to a SKF in terms of efficiency and accuracy.


\vspace{-.2cm}
\section{Locally concentrated hybrid state-space model}\label{section:model}
A spatial-temporal multi-modal LDS may be represented as
\begin{align*}
    \begin{bmatrix}
    \bm{x}_n{(1)}\\ \vdots \\\bm{x}_n {(d)}
    \end{bmatrix}&=
    \begin{bmatrix}
    \bm{A}_{s_{1,1}}(1,1)& \dots &\bm{A}_{s_{1,d}}(1,d)\\
    \bm{A}_{s_{2,1}}{(2,1)}&\dots &\bm{A}_{s_{2,d}}(2,d)\\
    \vdots & \ddots & \vdots \\
    \bm{A}_{s_{d,1}}(d,1)& \dots &\bm{A}_{s_{d,d}}(d,d)\\
    \end{bmatrix}
    \begin{bmatrix}\bm{x}_{n-1}{(1)}\\ \vdots \\\bm{x}_{n-1} {(d)}
    \end{bmatrix}
   \\
   &\quad \quad + \bm{\nu}_n,\\
    \bm{y}_n &= \bm{H}_n\bm{x}_n +\bm{\omega}_n.
\end{align*}
In the above, the state vector is $\bm{x}_n{(i)}$, where subscript $n$ and $i$ represent the time and the state index, respectively, $d$ is the dimension of the state, $\bm{A}$ is the evolution operator, and $s_{i,j}$ refers to a hidden switching random variable that determines the correlation between dynamic evolution of pixel $i$ with pixel $j$. The given information are $\bm{y}_n$, the measurement, $\bm{H}_n$, the measurement operator, and $\bm{\nu}_n \sim N(0,\bm{Q}_n)$, $\bm{\omega}_n \sim N(0,\bm{R}_n)$ [$N(\bm{\mu},\bm{\Sigma})$ refers to the Gaussian distribution with mean and covariance $\bm{\mu},\bm{\Sigma}$] are the evolution and measurement noise.

When the observed state variable is a function of variables that are locally concentrated in space, the evolution model in the neighboring pixels depend only on a single hidden random variable. It is assumed that each pixel in the image is correlated with its neighboring pixels with distance of at most $r$ pixels (which means $\bm{A}_{s_{i,j}}{(i,j)} = 0 \text{ if } |i-j|>r$). The state dynamic equation is then
\begin{align*}
    \begin{bmatrix}
    \bm{x}^{(1)}_n\\ \vdots\\\bm{x}^{(B)}_n
    \end{bmatrix}&=
    \begin{bmatrix}
    \bm{A}_{s_{1}}^{(1)}&...&0\\
    \bm{A}_{s_{2}}^{(2,1)}&...&0\\
    \vdots & \ddots & \vdots\\
    0& \dots &\bm{A}_{s_{B}}^{(B)}\\
    \end{bmatrix}
    \begin{bmatrix}\bm{x}^{(1)}_{n-1}\\ \vdots \\\bm{x}^{(B)}_{n-1}
    \end{bmatrix} + \bm{\nu}_n,
\end{align*}
where $B$ is the total number of patches, $\bm{x}_n^{(j)}$ refers to the pixels at time $n$ in the $j^{th}$ patch that follow the same dynamic behavior, $\bm{A}_{s}^{(i)}$ is the evolution model for patch $i$, and $\bm{A}_s^{(i,j)}$ is the correlating term between the dynamic model of patch $i$ with its neighboring patch $j\in\mathcal{B}_i$ where $\mathcal{B}_i$ is the set of patches neighboring of patch $i$.

\begin{figure*}
      \centering
  \centerline{\includegraphics[height=5cm]{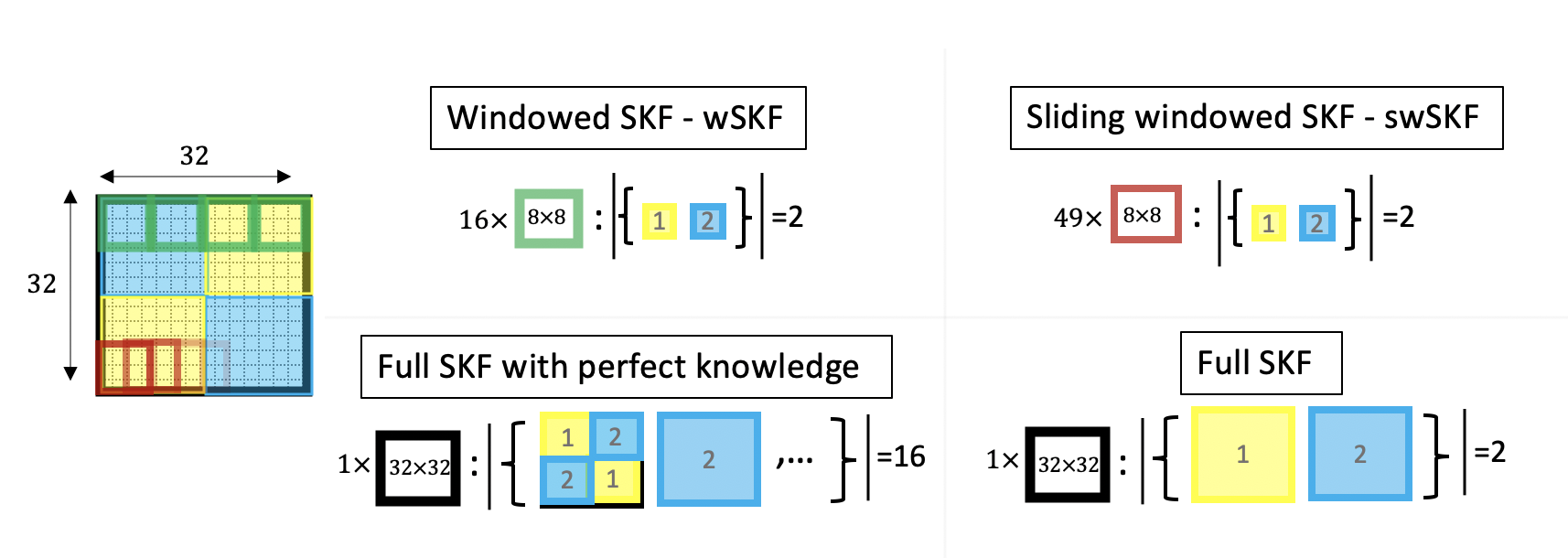}}
    \caption{Patch-based filtering vs.\ full filtering for the simulation setup; the dynamics of each quarter switch based on a hidden random variable between two modes (blue and yellow). The variance of the two modes is known. The full SKF with perfect knowledge has access to the spatial extent of the modes (quarter regions in this example), while the full SKF and (s)wSKF are equally ignorant of mode spatial content. }
    \label{fig:demo}
\end{figure*}

\vspace{-.2cm}
\section{Patch-based estimation}
\label{sec:pSKF}
For an LDS, the optimal Bayesian estimator is the Kalman filter (KF). In this section, we first review the KF/SKF procedure, as they are the building blocks of the proposed algorithm. Patch estimators using multiple low-dimensional SKFs are then proposed for efficient estimation of a spatial-temporal, locally concentrated evolving dynamic system.

\subsection{Kalman filter}\label{sec:kf}
With $\bm{y}_1^n$ defined as the set of all samples $\bm{y}_1,\bm{y}_2,...,\bm{y}_n$, the initial state is $\bm{x}_0\sim N(\bm{x}_{0|0},\bm{P}_{0|0})$, $\bm{x}_{n|n}=\E[\bm{x}_n|\bm{y}_1^n]$ and $\bm{P}_{n|n}=\C(\bm{x}_n|\bm{y}_1^n)$ are the conditional mean and covariance, and $L_n=p(\bm{y}_n|\bm{y}_1^{n-1})$ the likelihood, one step of the KF is
\begin{align}
    (\bm{x}_{n|n},\bm{P}_{n|n},L_n)&=Filter(\bm{A}_n,\bm{H}_n,\bm{x}_{n-1|n-1},\bm{P}_{n-1|n-1},\nonumber\\
    & \quad \quad\bm{Q}_n,\bm{R}_n,\bm{y}_1^n), \label{equation:filter}
\end{align}
which involves the following two recursive steps:\\
\textbf{Step 1: Time update equations}
\begin{align*}
\bm{x}_{n|n-1} &= \bm{A}_n \bm{x}_{n-1|n-1},\\
\bm{P}_{n|n-1}&= \bm{A}_n \bm{P}_{n-1|n-1}\bm{A}_n^T + \bm{Q}_n;
\end{align*}
\textbf{Step 2: Measurement Update equations}
\begin{align*}
\bm{e}_n &= \bm{y}_n - \bm{H}_n \bm{x}_{n|n-1}, \\
\bm{B}_n &:= \bm{H}_n \bm{P}_{n|n-1} \bm{H}_n^T + \bm{R}_n,\\
\bm{K}_n &= \bm{P}_{n|n-1} \bm{H}_n^T \bm{B}_n^{-1},\\
L_n &= N(\bm{e}_n; \bm{0}, \bm{B}_n),\\
\bm{x}_{n|n} &= \bm{x}_{n|n-1} + \bm{K}_n \bm{e}_n,\\
\bm{P}_{n|n} &= (\bm{I} - \bm{K}_n \bm{H}_n) \bm{P}_{n|n-1}.
\end{align*}
For an SLDS, $\bm{A}_n$ and $\bm{Q}_n$ can change with time and they must be detected as a component of the SKF using the obtained likelihoods ${L}_n$ using a Bayesian approach \cite{skf}.
\subsection{Patch estimator}\label{sec:patchskf}

The state-space equations for patch $i$  may be written as
\begin{align*}
    \bm{x}_n^{(i)} &= \bm{A}^{(i)}\bm{x}_{n-1}^{(i)}+ \sum_{j \in \mathcal{B}_i}\bm{A}^{(i,j)}\bm{x}_{n-1}^{(j)}+\bm{\nu}_n^{(i)},\\
    \bm{\Gamma}_i \bm{y}_n &= \bm{\Gamma}_i \bm{H} \bm{x}_n + \bm{\Gamma}_i \bm{\omega}_n,
\end{align*}
where $\bm{A}^{(i)}$, $\bm{A}^{(i,j)}$, and $\bm{Q}^{(i)}$ are functions of a hidden switching random variable corresponding to patch $i$ and neighboring patch $j$, and $\bm{\Gamma}_i$ is an operator applied to the measurements in order to find the localized measurements for process $i$. For localization, it is sufficient to have $\bm{\Gamma}_i\bm{H} \in \mathcal{T}= \begin{bmatrix}
    0,...,0\\
    0,\mathbf{\Theta^{(i)}},0\\
    0,...,0
    \end{bmatrix}$, if such $\bm{\Gamma_i},\bm{\Theta}_i$ exist. Otherwise, one must solve the constrained minimization problem $\argmin_{\bm{\Gamma}_i}{\|\bm{\Gamma}_i \bm{y}_n - \bm{\Gamma}_i \bm{H} \bm{x}_{n}\|_2^2}$ s.t. $\bm{\Gamma}_iH \in \mathcal{T}$. A special case is when local measurements are calculated for each pixel. In this case, the minimization problem to obtain local measurements is equivalent to solving an inverse problem, and the patch estimator may then be applied to the inverse problem's solution for denoising.

The estimation of a locally concentrated SLDS when the modes' spatial extent is unknown can be formulated as the following optimization problem for the set of possible dynamic models, patch sizes, shapes, and state variables:
 \begin{align}
     \argmin_{\bm{x}_n,s_i,\bm{U}_{i},\bm{U}_{j|i},B}&L(\bm{x}_n,s_i,\bm{U}_{i},\bm{U}_{j|i},B),\label{eq:opt0}
 \end{align}
 \vspace*{-1.5em}
 \begin{align}
     \label{eq:L}
     L&=\sum_{i=1}^B\sum_{j\in \mathcal{B}_i} \left\|\bm{U}_i \bm{x}_n^{(i)}-\bm{F}(s_i,\bm{U}_i {\bm{x}}_{n-1}) \right.  \\
     & \quad \quad \quad \left . -\bm{F}(s_i,\bm{U}_{j|i}\hat{\bm{x}}_{n-1}) \right\|_{(\bm{Q}_n^{(i)})^{-1}}^2 + \left \| \bm{H}\bm{x}_n-\bm{y}_n \right\| _{\bm{R}_n^{-1}}^2, \nonumber
\end{align}
such that $\bm{F}(s,\bm{b}) = {\bm{A}(s,\bm{b})}\bm{b}$ applies the linear operator ${\bm{A}(s,\bm{b})}$ to pixels in patch $\bm{b}$, $\bm{U}_i$ refers to the operator selecting the $i^{th}$ patch such that patches are disjoint and their intersection forms the whole image, and $\bm{U}_{j|i}$ selects the patches correlated with the $i^{th}$ patch. The loss function \eqref{eq:L} may be (approximately) decoupled when $\left \| \bm{H}\bm{x}_n-\bm{y}_n \right\| _{\bm{R}_n^{-1}}^2$ is replaced by $\sum_{i=1}^B\left\| \bm{\Theta}_i(\bm{U}_i)\bm{x}_n^{(i)}-\bm{\Gamma}_i(\bm{U}_i)\bm{y}_n \right\|_{\Delta_i[(\bm{\Gamma}_i\bm{R}_n\bm{\Gamma}_i^T)^{-1}]}^2$, where the matrix $\bm{\Gamma}_i$ is full rank and operator $\Delta_i$ selects the rows and columns of the matrix that correspond to the pixels in patch $i$. For now, we assume the matrix $\bm{H}$ to be invertible, so $\bm{\Gamma}_i$ always exists and is full rank. Solving the optimization \eqref{eq:opt0} for a general measurement operator is left for future work. 

Once the patches' structures are known, the optimization problem decouples into a minimization for each patch with $\argmin_{s_i,\bm{x}_n^{(i)}}L_i$ such that
\begin{align} \label{eq:opt2}
L_i&=\sum_{j\in \mathcal{B}_i} \left \|\bm{U}_i \bm{x}_n^{(i)}-\bm{F}(s_i,\bm{U}_i {\bm{x}}_{n-1})\right. \\
&\left.\quad \quad -\bm{F}(s_i,\bm{U}_{j|i}\hat{\bm{x}}_{n-1}) \right\|_{(\bm{Q}_n^{(i)})^{-1}}^2 \nonumber \\
& \quad\quad\quad\quad + \left \| \bm{\Theta}_i\bm{x}_n^{(i)}-\bm{\Gamma}_i\bm{y}_n \right\|_{\Delta_i[(\bm{\Gamma}_i\bm{R}_n\bm{\Gamma}_i^T)^{-1}]}^2\nonumber,
\end{align}
which is equivalent to running a SKF for the patch $i$. The additional terms corresponding to the neighboring patches are considered as inputs to the local state space equations.

Solving the optimization problem \eqref{eq:opt0} is an NP/hard problem since the loss function needs to be calculated for each possible patch size, shape, and location. Restricting the class of possible patch shapes to specific shapes and sizes reduces the computational burden. In practice, given the set of dynamic evolution models and $r$ (the correlation length), the image may be divided into windows of size $2r\times 2r$ and run the SKF for each window; we refer to this approach as the windowed SKF (wSKF). To reduce the effect of decoupling at window boundary points, it is also possible to use windows of size $(\alpha+2r)\times (\alpha+2r)$, slide the window over the whole image, and estimate each pixel using the centered window; we refer to this approach as the sliding windowed SKF (swSKF). The swSKF helps to reduce the boundary effects of windowing. It is notable that the window size must be greater than $2r\times 2r$ and that $\alpha>0$ must be chosen such that the windows are not too small (overfitting) or large (underfitting).

The decoupled estimation covariance will be larger compared the full SKF with knowledge of the mode spatial extent, since some information propagates through the image when the patches are dependent. Since running a SKF for every possible mode spatial extent is intractable, patch-based processing will have some information loss.

\begin{algorithm}
\SetAlgoLined
\KwResult{$\hat{\bm{x}}_1,\hat{\bm{x}}_2,...,\hat{\bm{x}}_n$ }
 \textbf{Input}: $\mathcal{F} = \{\bm{F}_1,\bm{F}_2,...,\bm{F}_l\}$ s.t. $\bm{F}_i\in\mathcal{F},
 \bm{H},\bm{R},\hat{\bm{x}}_0,\hat{\bm{P}}_0,r$,\\
 $\mathcal{Q} = \{\bm{Q}_1,...,\bm{Q}_l\}$ s.t. $\bm{Q}_i\in \mathcal{Q}$\\
 \For{$n=1:T$}{
  \For{$i\in \{1,...,B\}$}{
  $\hat{\bm{x}}_{n}^{(i)} = SKF(\mathcal{F},\mathcal{Q},\bm{\Gamma}_i \bm{y}_n,\bm{R},\bm{x}_0^{(i)},\bm{P}_0^{(i)})$
  }$\hat{\bm{x}}_n = [\hat{\bm{x}}_n;\hat{\bm{x}}_n^{(i)}]$}
 \caption{Patch-based estimation}
 \end{algorithm}
\begin{figure*}[h]
\subfloat[]{
  \includegraphics[width=.45\linewidth]{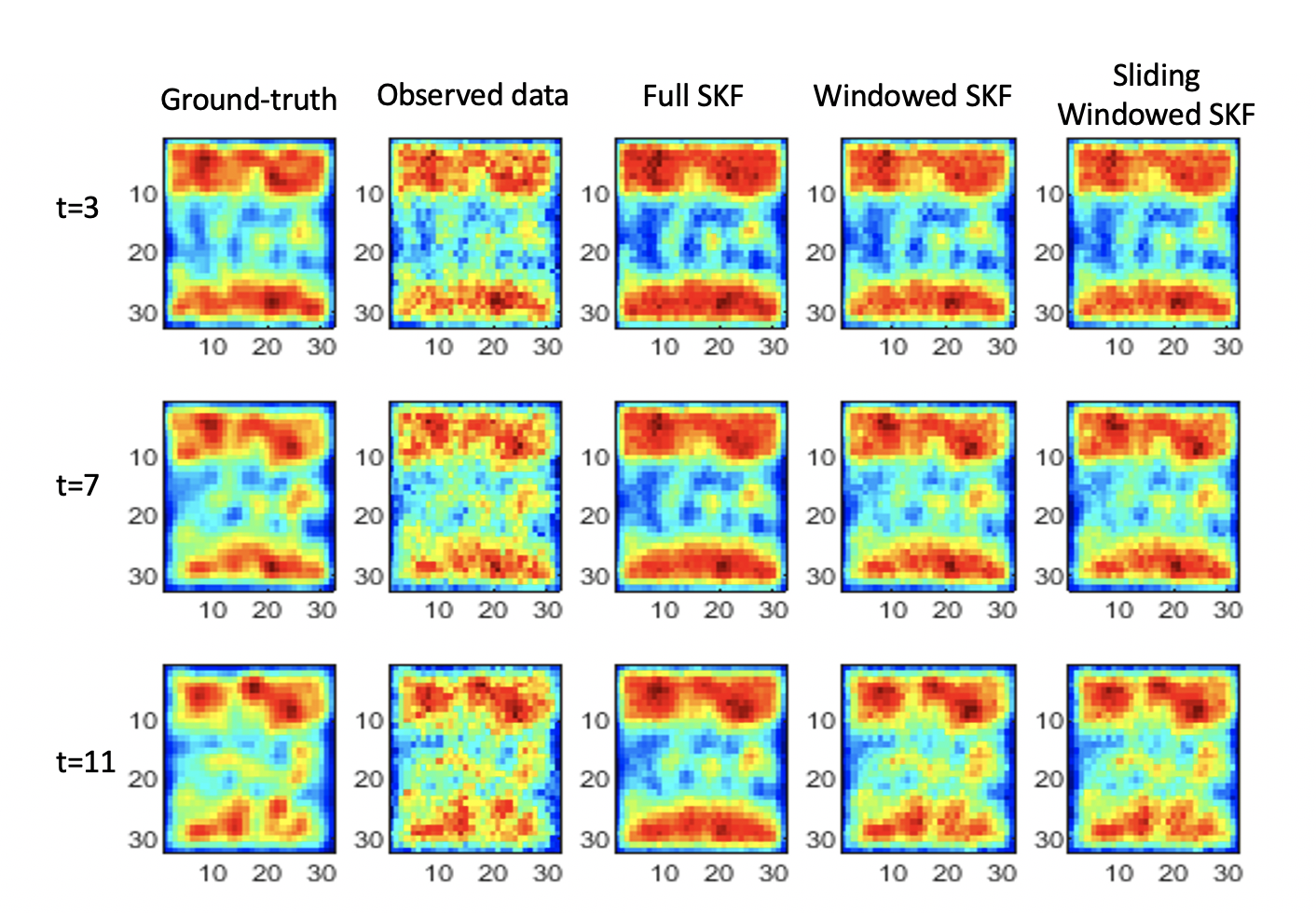}
  \label{fig:cloud}
  }
\subfloat[]{
    \includegraphics[width=.4\linewidth]{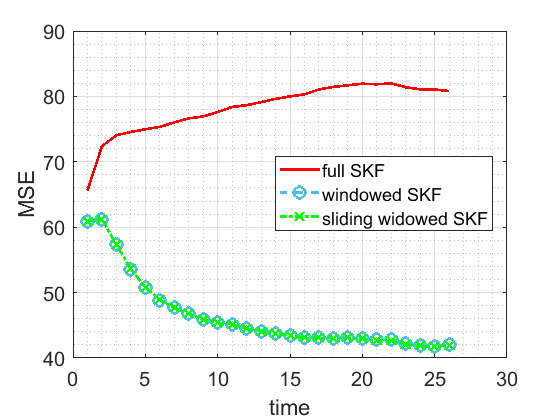}
    \label{fig:MSE}
    }
\caption{\protect \subref{fig:cloud} Realizations of the ground truth dynamical cloud images and the measurements, as well as the estimates using different filters. Visually, it is obvious that the (s)wSKF performs better than the full SKF with the same amount of information and with far fewer computations. \protect \subref{fig:MSE} The estimation MSE for the filters. The (s)wSKF has far less MSE compared to the full SKF with the same amount of information.}
\label{fig:all}
\end{figure*}
\vspace{-.2cm}
 \section{Computational cost} \label{section:comp}
Consider the estimation of a $\sqrt{d}\times \sqrt{d}$ image with $K$ non-overlapping processes that switch between $l$ modes over time. Thus, an SKF that has perfect knowledge of mode spatial extent must consider $K^l$ different modes at each time step with computational complexity $\mathcal{O}(K^ld^3)$. On the other hand, decoupling the image into $K$ windows where each window switches between $l$ modes only requires $\mathcal{O}(Klr_w^3)$ where $\sqrt{r_w}\times\sqrt{r_w}$ is the window size. (For an illustration, see Fig.\ref{fig:demo}). Similarly, running an swSKF requires $\mathcal{O}(K'lr_w^3)$, where $K'$ is the number of sliding windows and $K'>K$. Thus, the computational requirement of the patch-based SKF is much smaller than that of the full SKF, even given perfect knowledge of the spatial extent of the modes, and this difference becomes more significant as $d$, $K$, and $l$ increase.
\vspace{-.3cm}
\section{Simulation results}\label{section:sim}



 The performance of the proposed patch-based estimator for denoising reconstructed video frames from tomographic measurements is studied in this section. A sequence of $32\times 32$ images are reconstructed in time using a nonlinear, locally switching dynamic model, and a set of measurement with an average SNR of 11dB is generated accordingly; these images are meant to represent the evolution of cloud density obtained by infrared imaging in which clear local concentration is known to exist. The data are generated such that the structures' elements have random movements with slow/fast velocities (movement velocities are $0.01, 0.94$ pixels per time step) in each quarter of the image (as in Fig.~\ref{fig:demo}), and the velocity can switch randomly in each quarter over time. 
 
 The goal is to denoise the set of tomographically reconstructed images. Because the system dynamics are complicated, we use a purely stochastic model with two evolution covariance matrices corresponding to slow and fast velocities such that the variances are roughly set to the velocity of movements in the two cases. Assuming knowledge of the variances only, we apply a bi-modal full SKF, wSKF, and swSKF with window size $8\times 8$ to the sequence of images. Fig.\ref{fig:cloud} shows realizations of the ground truth, noisy reconstructed image, and the estimates for three representative time steps. The full SKF clearly cannot recover the ground truth details as well as the wSKF/swSKF. This conclusion is even more pronounced in Fig.\ref{fig:MSE}, where the error for each filter is averaged over 100 realizations and the full SKF MSE using the bi-modal model tends to diverge since the model used is mismatched. It is also notable that the swSKF error is slightly smaller compared to the wSKF. The (s)wSKF can find the estimates without the knowledge of the spatial extent of the modes and by detecting it instead while requiring fewer computations using the locality assumption. The run time of the filters are presented in the table below.
\begin{center}\label{table:1}
 \begin{tabular}{|c |c |c| c|}
 \hline
  & full SKF &  wSKF &   swSKF \\ [0.5ex]
  \hline
  Run time(sec) & 68.52 & 1.26& 4.18 \\ [0.5ex]
 \hline
 \end{tabular}
 \end{center}
\vspace{-.3cm}
\section{Conclusion}
A patch-based filtering framework is proposed for the estimation of image sequences governed by locally concentrated dynamics and shown to have superior performance with respect to the full filter in terms of computation and accuracy, when provided perfect/partial information about the evolution model. Solving the reconstruction problem for a general measurement operator, fusing the local estimates to obtain global estimates, and online learning of the evolution models/statistics of the patches locally are future work.

\bibliography{mycite}{}
\bibliographystyle{IEEEbib}

\end{document}